\documentclass[sigconf,preprint]{acmart}

\usepackage[all]{nowidow}

\begin{document}

\author{Tobias Pfandzelter}
\affiliation{%
    \institution{TU Berlin \& ECDF}
    \department{Mobile Cloud Computing}
    \city{Berlin}
    \country{Germany}}
\email{tp@mcc.tu-berlin.de}
\author{Jonathan Hasenburg}
\affiliation{%
    \institution{TU Berlin \& ECDF}
    \department{Mobile Cloud Computing}
    \city{Berlin}
    \country{Germany}}
\email{jh@mcc.tu-berlin.de}
\author{David Bermbach}
\affiliation{%
    \institution{TU Berlin \& ECDF}
    \department{Mobile Cloud Computing}
    \city{Berlin}
    \country{Germany}}
\email{db@mcc.tu-berlin.de}

\title{Towards a Computing Platform for the LEO Edge}

\acmYear{2021}\copyrightyear{2021}
\setcopyright{acmlicensed}
\acmConference[EdgeSys '21]{4th International Workshop on Edge Systems, Analytics and Networking}{April 26, 2021}{Online, United Kingdom}
\acmBooktitle{4th International Workshop on Edge Systems, Analytics and Networking (EdgeSys '21), April 26, 2021, Online, United Kingdom}
\acmPrice{15.00}
\acmDOI{10.1145/3434770.3459736}
\acmISBN{978-1-4503-8291-5/21/04}


\begin{abstract}
    The new space race is heating up as private companies such as SpaceX and Amazon are building large satellite constellations in low-earth orbit (LEO) to provide global broadband internet access.
    As the number of subscribers connected to this access network grows, it becomes necessary to investigate if and how edge computing concepts can be applied to LEO satellite networks.

    In this paper, we discuss the unique characteristics of the LEO edge and analyze the suitability of three organization paradigms for applications considering  developer requirements.
    We conclude that the serverless approach is the most promising solution, opening up the field for future research.
\end{abstract}


\maketitle

\section{Introduction}
\label{sec:introduction}

Private Internet and aerospace companies are currently building the largest satellite constellations in existence:
SpaceX, Amazon, and Telesat -- the so-called ``New Space'' companies -- are deploying or planning to launch tens of thousands of satellites into low Earth orbit (LEO) to provide global high-speed Internet access from space~\cite{Pultarova2015-ml}.
SpaceX's Starlink constellation has already entered a public beta phase for subscribers in North America and the United Kingdom with more than 1,500 satellites in use~\cite{Sheetz_undated-mc}.

Beyond enabling Internet access for underserved regions such as rural areas, planes, or cargo and passenger ships, these new satellite access networks can be a viable option for connecting all kinds of edge devices, even if terrestrial access is readily available.
High-bandwidth, low-latency connections to client ground stations and via inter-satellite links (ISL) enable direct low-latency routing between any two ground stations;
thus, many argue that satellite Internet will see broad adoption in the future~\cite{Bhattacherjee2018-vc,Bhattacherjee2019-jz,Handley2018-ay}.

Still, sending data from all end devices to a centralized location for processing can put a substantial strain on the satellite network.
Especially for applications that transfer large amounts of data or require low latency event processing, this can quickly become a problem~\cite{Pfandzelter2019-so}.
The same issue applies to terrestrial networks for which edge computing, a computing paradigm that uses compute resources at the edge of the network in direct proximity to end users and devices, has been proposed as a promising solution~\cite{Chandra2013-yf,paper_bermbach_fog_vision}.
Applications can use these resources through edge platforms that abstract from the geo-distributed server deployment and resource heterogeneity.
These platforms are based on different organization paradigms for applications (OPA), e.g., VMs, containers, or serverless functions.

Only recently, it has been proposed to also use computing resources at the LEO edge to facilitate novel applications serving a global user base~\cite{Bhattacherjee2020-kr}.
To realize this vision, however, edge platforms must consider the unique characteristics of the LEO edge while still satisfying general requirements of application developers.
As a first step towards this goal, we must consider the core of such a platform, the OPA, and analyze different options in light of the challenges at the LEO edge.
We therefore make the following contributions:
\begin{enumerate}
    \item We discuss the unique characteristics of the LEO edge (Section~\ref{sec:infrastructure}).
    \item We derive requirements for building LEO edge applications from a developer perspective (Section~\ref{sec:requirements}).
    \item We analyze the suitability of three OPAs with regards to how they might satisfy developer requirements considering LEO edge characteristics (Section~\ref{sec:evaluation}).
\end{enumerate}
\section{Background \& Related Work}
\label{sec:background}

In this section, we briefly introduce and describe the state of the art for large LEO satellite communication networks, edge computing, and OPAs.
Furthermore, we also provide an overview of existing LEO edge computing research.

\subsection{LEO Communication Networks}

Satellite-backed Internet access has been available for decades with geostationary satellites at altitudes of 35,000km.
The high altitude combined with the requirement to orbit above the equator, however, result in high access latency for consumers which renders satellite Internet inviable for many use cases~\cite{Clarke1945-qb,Iida2000-il}.

Now, a new generation of Internet satellites is being developed by companies such as SpaceX, Amazon, and Telesat.
Thousands of these satellites are being deployed in large constellations at altitudes of below 600km, i.e., in the LEO.
Thus, satellites continuously orbit the globe and cover large ground distances in short periods of time.
Each satellite is equipped with radio transmitters and receivers to connect to ground stations.
As ground station equipment is small enough for use in family homes or airplanes, each satellite serves many user terminals concurrently~\cite{Bhattacherjee2018-vc,Bhattacherjee2019-jz,Handley2018-ay,Handley2019-ce}.

\begin{figure}
    \centering
    \includegraphics[width=0.75\columnwidth]{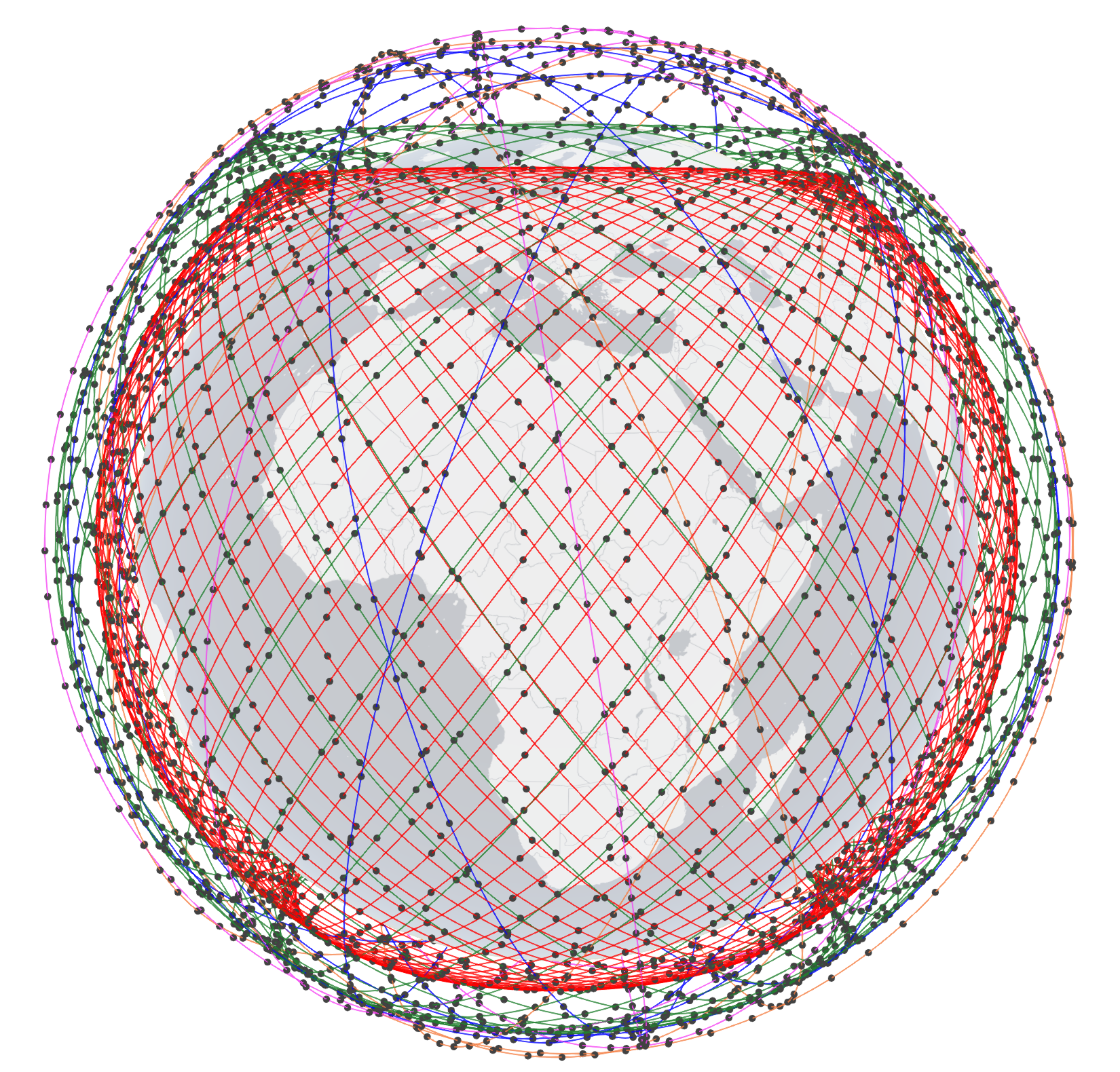}
    \caption{Overview of a LEO satellite constellation comprising different shells. Shown here is the proposed first phase of the Starlink deployment with five shells of 1,584 satellites at 550km, 1,600 at 1110km, 400 at 1130km, 375 at 1275km, and 450 at 1325km altitude~\cite{Kassing2020-yc}.}
    \Description[Overview of a LEO satellite constellation]{LEO satellite constellations comprise thousands of satellites arranged in shells of evenly-spaced orbits.}
    \label{fig:constellation}
\end{figure}

A complete LEO satellite constellation comprises different shells of satellites.
Each shell is a collection of orbital planes with the same orbital parameters equally distributed across the earth, with each plane comprising a number of equally spaced satellites.
We show such a constellation in Figure~\ref{fig:constellation}.

Given the low altitude, a single satellite has a relatively small cone of coverage yet it must be connected to some form of uplink to provide Internet access.
If no ground station uplink is available, e.g., because a satellite currently crosses an ocean, satellites may connect to adjacent satellites via ISLs.
This way, satellites can acquire Internet access even if no ground station is within their field of view.
Satellites also use ISLs for providing low latency broadband access since light propagates faster in a vacuum than in fiber.
This makes satellite based Internet an attractive alternative for many Internet subscribers~\cite{Khan2015-wf,Bhattacherjee2018-vc,mustk-lf}.

A main driver of the ``new space race'' are decreasing satellite launch cost due to the development of reusable rockets such as SpaceX's Falcon 9 launch vehicle~\cite{Jones2018-ct}.
Yet building a LEO satellite constellation from scratch still requires large upfront investments.
Furthermore, regulatory challenges prove to be another market entry barrier.
Despite its size, LEO is a scarce resource as satellite collisions have to be averted.
To this end, satellites are also equipped with the capability to dodge obstacles.
Additionally, companies must also register the usage of radio frequencies for the communication between ground stations and satellites.
These factors mean that any change to the constellations requires a substantial lead time~\cite{Bhattacherjee2018-vc,Sheetz_undated-ea}.

\subsection{Edge Computing}

The demand for processing data close to its origin led to an increased popularity of the edge computing paradigm in research and industry~\cite{Chandra2013-yf,paper_bermbach_fog_vision}.
The main idea behind edge computing is to embed computing resources into the edge of the network, i.e., close to clients.
Compared to cloud computing, resources are thus available with low latency and bandwidth costs.
Preventing data from being transmitted to the cloud can also reduce privacy and security risks~\cite{paper_pallas_fog4privacy}.

Typically, edge computing infrastructures comprise a multitude of geo-distributed nodes with different compute, storage, and network capabilities.
Using such a heterogeneous infrastructure is significantly more difficult than the ease-of-adoption developers are used to from the cloud:
Applications need to consider geo-distribution, horizontal scalability, and network availability for fully leveraging edge resources.
To remedy this, a number of edge computing platforms have been proposed to abstract from the underlying infrastructure.
To this end, they employ service offloading techniques, data geo-replication, or resource sharing~\cite{paper_bermbach_fog_vision,Pfandzelter2020-kw,paper_hasenburg_towards_fbase,techreport_hasenburg_2019,Satyanarayanan2009-pc}.

\begin{figure}
    \centering
    \includegraphics[width=\columnwidth]{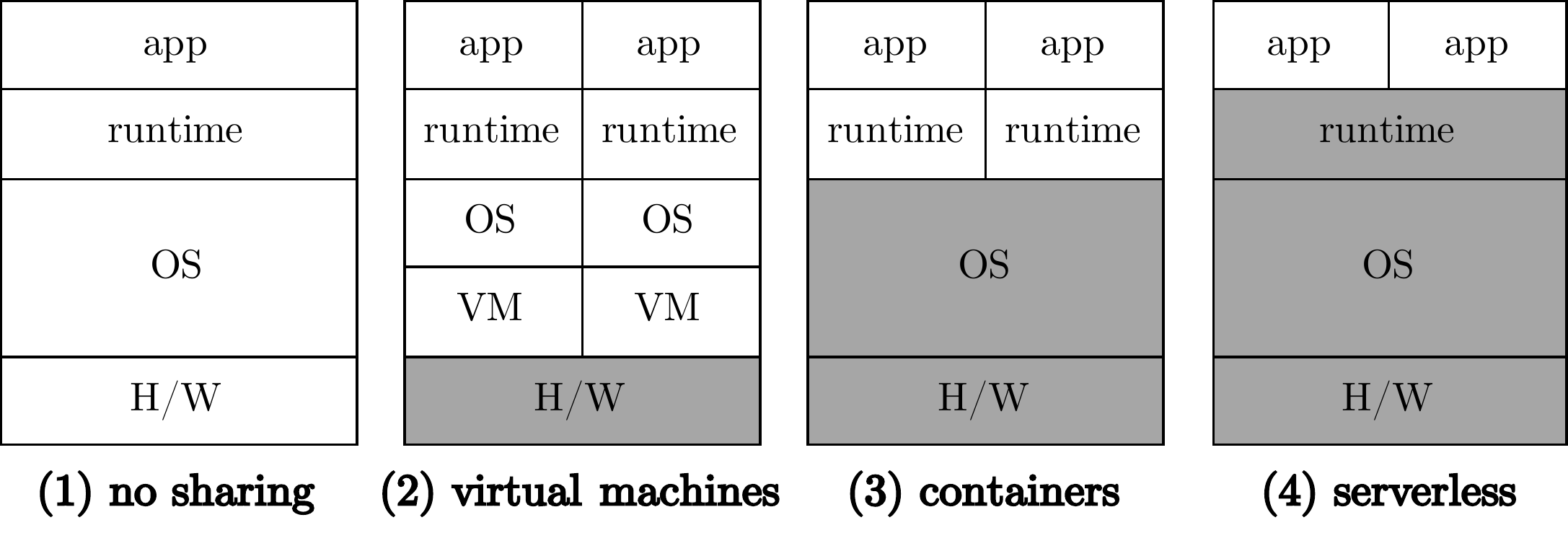}
    \caption{OPAs by levels of abstraction~\cite{Hendrickson2016-pw}: while less abstraction offers more freedom and flexibility, more abstraction simplifies writing applications and shifts execution control to the platform which often increases efficiency.}
    \Description[OPA by levels of abstraction]{while less abstraction offers more freedom and flexibility, more abstraction simplifies writing applications and shifts execution control to the platform which often increases efficiency.}
    \label{fig:opa}
\end{figure}

Edge computing platforms are usually developed around a specific OPA.
In Figure~\ref{fig:opa}, we show that the three main paradigms that have initially evolved in the context of cloud computing offer different levels of abstractions.

\paragraph{Virtual Machines}

Virtual machines (VM) are virtualized servers that run operating systems which have often been slightly modified for VM usage to increase performance.
Multiple virtual machines can share a common physical machine, hence a single server can be made to look like many smaller machines.
VMs can be used just like regular servers and host long-running applications, often with multiple connected services sharing a virtual host~\cite{Chandra2013-yf}.

\paragraph{Containers}

Containers encapsulate individual processes instead of full servers.
Each service can be packaged into a container image together with its dependencies and deployed alongside other containers on a common host.
Containers have the benefit that no entire operating system has to be managed by application developers but rather only the actual service and its dependencies.
This also reduces the application footprint, making container migration easier than VM migration.
Containers do not change the way processes are executed, so they can also be used for long-running applications~\cite{Morabito2018-ip}.

\paragraph{Serverless Functions}

With serverless functions, only the actual code is deployed while the runtime is provided by the platform.
Serverless functions are short-lived, i.e., they exist only for a single invocation.
As such, they cannot maintain state yet provide scalability as additional, parallel function instances can be launched to distribute requests.
Conversely, function instances only consume resources during their execution~\cite{Pfandzelter2020-kw}.
Nevertheless, even serverless applications must maintain state somehow; to that end, serverless databases, often following the NoSQL paradigm, are employed~\cite{paper_hasenburg_towards_fbase,techreport_hasenburg_2019}.\\

The distinction between these paradigms is not always clear in their implementation.
For example, a serverless function platform may use containers as runtimes for their functions~\cite{Pfandzelter2020-kw}.
Similarly, a container orchestrator can run containers on virtual machines.
A fourth paradigm, the \emph{unikernel}, is currently on the rise.
As with containers, unikernels package applications and their runtimes, yet they also add a library operating system that obviates the need for a shared OS~\cite{Madhavapeddy2013-lx}.
We do not consider them separately in this paper as they combine concepts of virtual machines and containers.

\subsection{Related Work}

The highly dynamic nature of satellite constellations and their limited capacity for computational power means that existing edge computing platforms are not yet ready for being applied to the LEO edge.
Based on this premise, Bhosale et al.~\cite{Bhosale2020-aa} propose \emph{Krios}, a new platform based on Kubernetes that manages stateless as well as stateful services within the LEO edge.
Through ground station servers that calculate satellite orbital positions, Krios is able to predict when a service has to be spawned in a different satellite in a ``just-ahead-of-time'' manner.
Krios is therefore able to provide services without downtime from a client perspective.
While this is an important first proposal, the authors decision for containers seems somewhat arbitrary as they skip the evaluation of different paradigms, as we do in this paper.

The feasibility study conducted by Bhattacherjee et al.~\cite{Bhattacherjee2020-kr} analyzes possible use cases and hand-off strategies, and explores the viability of establishing computing resources in space.
The authors specifically mention content delivery networks, a use case we have also investigated in more detail in a previous study~\cite{pfandzelter2020edge}, augmented reality, ``Tactile Internet'' applications, multi-user interaction, and space-native data processing as possible use-cases.
While such applications are also supported by terrestrial edge computing, the authors note that adding compute capabilities to LEO satellites can help to bridge the digital divide by bringing them to underserved areas as well.
The presented hand-off strategy is based on the concept of ``virtual stationarity'' to abstract from the dynamic nature of the satellites and make the server appear as a single entity to the clients.
For this, a heuristic picks a satellite server from a set of options with a near-optimal latency that is also visible to the client for the longest time, thus reducing the amount of required hand-offs.
The next logical step beyond this initial feasibility study and effective server selection strategy is to build a platform that enables real application to take advantage of LEO edge computing.
As a prerequisite, however, we first have to analyze different OPAs with regard to their suitability for the LEO edge.

\section{LEO Edge Characteristics}
\label{sec:infrastructure}

In this section, we discuss the unique characteristics of the LEO edge.
At the time of writing, LEO satellites can only be used to connect to the Internet.
Computing infrastructure, e.g., in the form of servers located at each individual satellite~\cite{Bhattacherjee2020-kr,Bhosale2020-aa}, is not yet available to the general public.
Thus, this discussion makes likely assumptions in some parts, especially concerning hardware capabilities.

\paragraph*{C1: Mobile Server Infrastructure}

The first thing to keep in mind with servers attached to satellites in a LEO constellation is that these satellites orbit the earth at high speeds.
For example, a satellite at an altitude of 550km must maintain a speed of 27,000km/h to maintain its orbit~\cite{Bhattacherjee2019-jz}.
Consequently, the servers also move at this speed.
For the static ground station equipment this means that they must frequently change their communication partner.

\paragraph*{C2: Same-model Servers}

Then, satellites in a constellation are mostly the same model.
The reason for this is that satellites orbit the earth continuously while the earth revolves beneath the satellite constellation.
Thus, each satellite eventually covers each part of the earth which means that using different kinds of satellites for different regions is not possible.
Subsequently, the servers must also be of the same model.
It can be possible to upgrade server capabilities over time as satellites reach the end of their lifetime, yet developing different versions can have a negative impact on development and production costs.
Hence, we believe that it is likely that hardware capabilities will be comparable across satellites until a new generation of satellites (then with again improved hardware) is launched.
We believe that it is unlikely that more than two or three hardware versions will be deployed in parallel.

\paragraph*{C3: Homogeneously Distributed Servers}

Due to their non-geosta\-tionary nature, satellites are also homogeneously distributed across the globe, with satellites evenly spaced across an orbit.
This means that each ground station has access to more or less the same amount of equally equipped satellites at all times.

\paragraph*{C4: Heterogeneous Demand}

Nevertheless, demand is of course not homogenous across earth.
Urban areas have a higher client density which increases resource demand compared to rural areas or oceans with a smaller client population.

\paragraph*{C5: Limited Compute Capabilities}

As a consequence of being deployed in space, satellite servers' capabilities must be limited.
The reason for this is that energy consumption and heat generation must be kept low for economical reasons.
Larger heat dissipation mechanisms, batteries, or solar arrays lead to higher weight and, subsequently, higher launch costs~\cite{Bhattacherjee2020-kr}.

\paragraph*{C6: No Physical Access}

Another effect of placing servers on satellites in LEO is that those servers cannot be accessed for maintenance.
Consequently, if a satellite or server fails, it remains failed and can only be de-orbited.

\paragraph*{C7: Fixed Server Capabilities}

Not being able to access individual servers directly also means that they cannot be upgraded.
Over the lifetime of a satellite, typically about 5 years~\cite{Sheetz_undated-dm}, the server capabilities and, with it, the total capability of the constellation of servers, remain fixed.

\paragraph*{C8: Fixed Number of Servers}

Horizontal scalability is also limited, as we can place servers only on satellites that are part of the constellation and the size of the constellation cannot be changed easily.
Launching and deploying additional satellites requires approval by governmental agencies and competing space Internet companies may lobby to limit constellation sizes, especially as LEO is a limited resource~\cite{Bhattacherjee2018-vc,Sheetz_undated-ea}.
\section{LEO Edge Application Requirements}
\label{sec:requirements}

In this section we derive requirements for building applications using a LEO edge platform from the perspective of a developer that is used to developing cloud applications.
Some of the listed requirements are not particular to the LEO edge but also apply to, for example, edge platforms that are designed for terrestrial networks.

\paragraph{R1: Deployment Close to Clients}

The main advantage of edge over cloud computing is that the services of an application can be deployed close to their users.
As such, if an application is deployed on a LEO edge platform, the platform should take care that individual services are placed on the parts of the infrastructure that are in proximity to their clients~\cite{Chandra2013-yf,paper_bermbach_fog_vision}.

\paragraph{R2: High Availability}

As with cloud computing, developers expect their application to be highly available in a LEO edge environment as well.
Consequently, a LEO edge platform needs to abstract from the widely distributed and heterogeneous underlying infrastructure to provide fault-tolerance~\cite{paper_bermbach_fog_vision,Morabito2018-ip}.

\paragraph{R3: Isolation}

A single LEO edge platform can be used to host a number of services from different developers.
Ideally, this multi-tenancy should be hidden from developers, as other services should not interfere with each other.
This comprises two dimensions: security and performance isolation.
From a security perspective, a service must neither be able to identify what other services are using the platform, nor should it be able to read other services' data~\cite{Morabito2018-ip,Satyanarayanan2009-pc}.
The performance of a service must also not be impacted by other services, i.e., no colocated service should degrade performance on the same machine for other services.
Performance degradation could, for example, occur when two such services share common hardware without sufficient measures to limit resource utilization~\cite{Jeyakumar2013-if}.

\paragraph{R4: Familiar, Open Technology Stack}

To simplify the transformation of cloud-client applications to cloud-edge-client applications, developers should be able to develop services for LEO edge  platforms using a familiar technology stack, i.e., one that can also be applied in the cloud.
This encompasses processor architectures (e.g., x86, x64, arm), operating systems (e.g., Linux, Windows), and programming languages (e.g., Java, Go, Python).
The need for using novel technologies may hinder the adoption of the platform given a steep learning curve and larger upfront investments~\cite{Morabito2018-ip,Satyanarayanan2009-pc}.

\paragraph{R5: Flexible Deployment}

Cloud computing's ``pay-as-you-go'' mod\-el has been a major factor in its success as it enables flexible service deployment without long term commitments.
A similar deployment model will likely increase adoption of a LEO edge platform.
Developers want the flexibility to create new applications and remove deprecated services while being charged only for the infrastructure they use~\cite{paper_bermbach_fog_vision,Bermbach2020-hg}.

\paragraph{R6: Elastic Scalability}

Another important feature of cloud computing is elastic scalability, i.e., that services can take advantage of additional infrastructure if demand increases and release infrastructure if it decreases~\cite{book_cloud_service_benchmarking}.
To this end, the cloud provides the illusion of infinite resources.
A LEO edge platform should provide a similar illusion and provide the same level of scalability so that services do not fail when demand increases.
We note that in edge computing or at the LEO edge this is especially difficult given that the distributed underlying infrastructure cannot as easily benefit from resource pooling~\cite{Morabito2018-ip,Pfandzelter2020-kw}.
\section{Suitability of OPA Options to LEO Edge Scenarios}
\label{sec:evaluation}

In this section, we analyze how different organization paradigms for applications (OPAs) can be used for building a LEO edge platform that satisfies the requirements presented in Section~\ref{sec:requirements} considering the unique characteristics of the LEO edge as discussed in Section~\ref{sec:infrastructure}.
For this analysis, we consider the following OPAs: virtual machines, containers, and serverless functions.

\paragraph{R1: Deployment Close to Clients}

To deploy a service close to clients, two steps are necessary:
First, client locations have to be identified, either upfront in a static manner or dynamically at runtime.
Second, infrastructure in proximity has to be identified and allocated so that the service can be deployed.

On the LEO edge, two factors make this a particularly difficult problem.
First, the LEO edge, by design, provides global coverage, so global clients have to be considered, whereas on the terrestrial edge a platform only covers a specific country or area.
Scheduling thus has to happen in a distributed manner rather than with a central server, as a centralized scheduler, whether static or dynamic, can easily become a bottleneck given the large amount of global clients.
Additionally, the combination of homogeneously deployed satellite servers (\emph{C3}) and heterogeneous client demand (\emph{C4}) must be taken into account.
This scheduling component is an orthogonal platform design question compared to choosing the right OPA, yet we note that the traditional centralized cluster orchestrator in use by container orchestration tools such as Kubernetes is a hindrance in this context.

Second, satellites are mobile (\emph{C1}).
Ground stations connect to their nearest satellite and expect their service to be available at that satellite.
As satellites orbit the earth, ground stations reconnect and applications have to either move across the LEO edge infrastructure to remain stationary in relation to the clients or be deployed across the entire infrastructure so that it is always accessible.
Virtual machines, containers, and serverless functions can all be moved between servers as required to provide this virtual stationarity, albeit with different complexity.
Both virtual machines and containerized services can be migrated live without downtime, yet the larger footprint of a virtual machine leads to a higher migration overhead.
For stateful applications, the overhead can quickly become significant when migrations occur frequently, e.g., as it is the case for large satellite constellations and fast-moving satellites.
Stateless serverless functions, on the other hand, benefit from their inherent concurrency.
Static function code can be preemptively propagated to nearby satellite servers and new instances can already be instantiated
Alternatively, functions could simultaneously be deployed on all satellite servers concurrently given the small footprint of function code.
They would not necessarily also have to be instantiated at all times but could rather start once the first request arrives (cold start).
The overhead of frequent cold starts may, on the other hand, be remedied by providing hints~\cite{paper_bermbach_faas_coldstarts} about client location in regards to the mobile satellites since their movement is highly predictable.
To manage state in a stateless serverless environment, additional stateful services such as database systems are needed.
Here, a shared serverless database infrastructure may be provided where techniques similar to VM and container migration can be employed, albeit with a smaller data footprint as only application state has to be transferred~\cite{techreport_hasenburg_2019,paper_hasenburg_towards_fbase}.

\paragraph{R2: High Availability}

To provide high availability in the presence of server or satellite failure, services need to be migrated to nearby servers since servers cannot easily be repaired or replaced (\emph{C5}).
This is problematic for stateful VMs and containers, as only a single instance can be live at one time and this instance needs to be reinstantiated on a new satellite to offload the service.
In case of satellite failures, a replicated live backup instance may even be required so no state is lost.
This increases service overhead, which is problematic on limited satellite servers (\emph{C6}).
In case of stateless VMs or containers, this is still possible but suffers from the same migration costs discussed above.
The serverless OPA is a better fit in this case as backup functions can be deployed on nearby satellites or across the entire constellation without state conflicts.
If state is managed by a shared serverless database, correct data backups or replication can be efficiently provided to the application.

\paragraph{R3: Isolation}

All OPAs provide some level of security isolation depending on their implementation, so performance isolation is a far more pressing issue on the limited hardware of a satellite server (\emph{C5}).
For each OPA, the available resources can be provisioned safely, i.e., that only the actually available resources are allocated.
On the other hand, over-provisioning has an economical benefit as not every service uses its allocated resources at all times.

A VM's resource allocation is static for its entire lifetime and since it hosts an entire operating system, its required resources are comparatively large.
This coarsely grained resource allocation is less efficient than the alternatives.
For containers, which host individual processes, resources do not have to be allocated but can rather be limited to prevent leaking into other services.
These limits can also be adapted over the lifetime of a container.
Depending on the implementation of the serverless execution environment, resources for serverless function instances may also be limited or statically allocated, although function lifetime is considerably shorter than that of containers or VMs.
Resource allocation can thus be maximized at all times without degrading service quality.

\paragraph{R4: Familiar, Open Technology Stack}

All three OPAs are widely used in the context of cloud computing and can thus be considered to reflect familiar technology stacks.
Nevertheless, we must also consider that some of the decisions made regarding the infrastructure of satellite servers may ``leak'' through the layers of abstraction.
For example, if a novel ``space-ready'' processor architecture should be employed to support the unique LEO environment (\emph{C5}), developers would be unable to deploy a VM with an operating system that does not support this architecture.
Containers provide a higher level of abstraction, so just the chosen language runtime needs to support this architecture.
Serverless functions provide the highest level of abstraction since the platform also provides the language runtime.

On the other hand, lower levels of abstraction, i.e., virtual machines, also offer more freedom when it comes to bringing one's own technology stack, so that it is easier to transfer familiar technologies from terrestrial edge and cloud to the LEO edge.

\paragraph{R5: Flexible Deployment}

Shipping new VM and container images or serverless function code to satellite servers is simple.
All three OPAs also provide the flexibility to quickly start and stop services.
Flexible deployment, however, also requires proper resource allocation.
As the servers' capabilities are limited (\emph{C7}) and scaling services arbitrarily far up or out to meet a growing demand is not an option (\emph{C8}), merely the illusion of infinite resource can be provided.
To that end, a market-based approach where the limited resources are provisioned for the highest bidder could be an option~\cite{Bermbach2020-hg}.
In combination with a pay-as-you-go model based on resources used in a specific time span, profit can be maximized.
The efficacy of this approaches improves with lower granularity of allocated resources, i.e., serverless functions are more efficient than containers and VMs.
Of course, offloading as proposed for the availability requirement (\emph{R3}) above is also an option to some extent.

\paragraph{R6: Elastic Scalability}

Given the heterogeneous service demand (\emph{C4}) in combination with a fixed amount (\emph{C8}) of homogeneously distributed satellite servers (\emph{C3}) that cannot be upgraded (\emph{C7}), off\-loading is required if the demand for one service exceeds the limited capabilities of a single server (\emph{C5}).
Individual service requests or even full services may need to be offloaded.
Request offloading is not possible given that only one copy of a stateful VM or container can be instantiated at a time, so the entire service may need to be moved to a nearby satellite server with enough capacity.
For stateless container or VM implementations, offloading of requests is possible yet results in high resource costs.
In contrast to service migrations due to satellite mobility, changes in demand can also not be predicted as easily.
Additionally, containers and especially VMs incur a large migration overhead.
On the other hand, the low latency ISLs make forwarding requests to nearby satellites easier.
In the case of serverless functions, this can be used to offload individual requests in case the local server has no more capacity.\\

Overall, the three OPAs we analyzed use different levels of abstraction, from VMs to serverless functions.
Our analysis shows that more abstraction gives greater control to the LEO edge computing platform, which is important to satisfy developer requirements in light of the unique characteristics of LEO edge infrastructure.
Especially mobility of servers is a characteristics that goes beyond what terrestrial edge computing platforms must usually support in practice.
We thus conclude that a high level of abstraction benefits upcoming LEO edge platforms and that the serverless paradigm will be the best choice for future work in this field.
Of course, all OPAs can be used, yet at different cost levels.
\section{Conclusion \& Future Work}
\label{sec:conclusion}

In this paper, we analyzed the suitability of three OPAs for upcoming LEO edge platforms.
For this, we discussed the unique characteristics of the LEO edge and derived requirements for building applications using such a platform from the perspective of a developer that is used to developing cloud applications.
We found that low performance overheads, flexible reconfiguration and mobility of application components, and strong support for multi-tenancy are essential for a LEO edge platform.
Based on this, we conclude that the serverless OPA is the best fit to fulfill these needs, as individual functions can provide efficient resource allocation even in constrained environments such as LEO satellite servers -- still the other OPAs can also be used but have additional costs and limitations.
Finally, functions provide concurrency out of the box, which simplifies mobility and flexible deployment; also multi-tenancy is possible through the high degree of resource sharing.
In future work, we plan to design and evaluate such a serverless platform for the LEO edge.

This paper should be seen as a foundation for new research on LEO edge platforms, as a range of open research questions arises from it:
For example, while virtual stationarity can be provided by anticipating satellite and earth movement, the concept of ``stationarity'' should be further explored since it can be achieved on a ground station, city, or even country level.
In this regard, the efficiency of different hand-off models should be investigated as well.
Handing off to the satellite that is closest to a specific ground station leads to the best access latency yet increases overhead.
More efficient techniques such as employed in~\cite{Bhattacherjee2020-kr} should be evaluated in realistic environments.
Another question is how LEO edge-based platforms can integrate with existing terrestrial cloud/edge/fog infrastructure -- can LEO edge platforms benefit from them?
Finally, testing and benchmarking LEO edge platforms is hard without access to the corresponding infrastructure.
To that end, we also propose to design testbeds for emulation or simulation of LEO edge platforms.

\begin{acks}
    Funded by the \grantsponsor{DFG}{Deutsche Forschungsgemeinschaft (DFG, German Research Foundation)}{https://www.dfg.de/en/} -- \grantnum{DFG}{415899119}.
\end{acks}

\balance

\bibliographystyle{ACM-Reference-Format}
\bibliography{bibliography}

\end{document}